\documentclass[preprint,prc,aps,superscriptaddress,nofootinbib,showpacs]{revtex4}
\usepackage[dvips]{graphicx}
\usepackage{amsfonts, color, float, bbm, amsmath}
\usepackage{amssymb}

\def\beq {\begin{equation}}
\def\eeq {\end{equation}}
\def\bea {\begin{eqnarray}}
\def\eea {\end{eqnarray}}

\def\ni {\noindent}

\def\s {\sigma}

\def\p {\pi}

\def\p {\mbox{\boldmath $\pi$}}

\begin{document}

\title{NUCLEAR FORCES and CHIRAL SYMMETRY\footnote{Talk given at The Third Asia-Pacific Conference 
on Few-Body Problems, Nakhon Ratchasima, Thailand, July 2005}}

\author{R. Higa}

\address{Thomas Jefferson National Accelerator Facility,\\
1200 Jefferson Avenue, Newport News, VA 23606, USA\\ 
E-mail: higa@jlab.org}

\author{\underline{M. R. Robilotta}}

\address{Instituto de F\'{\i}sica, Universidade de S\~{a}o Paulo,\\
C.P. 66318, 05315-970, S\~{a}o Paulo, SP, Brazil\\
E-mail: robilotta@if.usp.br}  

\author{C. A. da Rocha}

\address{N\'ucleo de Pesquisa em Bioengenharia, Universiade S\~{a}o Judas Tadeu,\\
Rua Taquari, 546, 03166-000, S\~{a}o Paulo, SP, Brazil\\
E-mail: carlos.rocha@usjt.br}

\begin{abstract}
We review the main achievements of the research programme for the study of nuclear forces 
in the framework of chiral symmetry and discuss some problems which are still open.

\end{abstract}

\maketitle

The research programme for the study of nuclear forces, based on the idea that long and 
medium range interactions are dominated by one and two-pion exchanges, was formulated
more that fifty years ago in a seminal paper by Taketani, Nakamura and Sasaki\cite{TNS}.
Nevertheless, only in the last fifteen years it has achieved its full strength, 
due to the consistent use of chiral symmetry.
This led to a considerable improvement in our knowledge of the basic mechanisms 
underlying nuclear interactions.
For a comprehensive discussion of the subject, the reader is directed to the recent
review produced by Epelbaum\cite{EE}.
Here we describe the main topics in which progress has been made towards clarifying dynamics 
and outline some problems which still remain open, in a perspective biased by the work
done by our group\cite{HRR}.

\vspace{1mm}

The works by Weinberg\cite{WNN} in the early nineties motivated the systematic use 
of chiral symmetry in the study of nuclear forces.
The rationale for this approach is the fact that nuclear interactions are dominated 
by low-energy processes involving the quarks $u$ and $d$, which have small masses.
This allows one to work with a two-flavor version of QCD and to treat these masses 
as a perturbation in a chiral symmetric massless lagrangian.
The procedure for the systematic inclusion of the effects associated with the quark masses 
is known as chiral perturbation theory (ChPT). 
In order to be able to perform chiral expansions, one uses a typical scale $q$, 
set by either pion four-momenta or nucleon three-momenta, such that $q<1$ GeV.

\vspace{1mm}

Chiral symmetry is especially suited for dealing with multipion processes.
Hence, in the case of the one-pion exchange potential $(OPEP)$, it becomes relevant only
when form factors are taken into account.
On the other hand, it is essential to the accurate description of the two-pion exchange
potential $(TPEP)$, which is closely related to the $\p N$ scattering amplitude.
In the sequence, we concentrate on this component of the force.

\vspace{1mm}

The leading contribution to the $TPEP$ is $O(q^2)$ and, at present, 
there are two independent expansions of the potential up to $O(q^4)$ in the literature, 
based on either heavy baryon\cite{HB} or covariant\cite{HRR} ChPT.
These results allow one to put the problem in perspective and note that 
the following aspects of the problem have been tamed:\\[1mm]
$\bullet$ Quite generally, asymptotic (large $r$) expressions for the potential have the status 
of theorems and are written as sums of chiral layers, with little model dependence.\\[1mm]
$\bullet$ The minimal realization of the symmetry is implemented by just pions and nucleons,
but realistic potentials require other degrees of freedom, either hidden within 
the-low energy constants (LECs) of effective lagrangians or represented as explicit deltas. \\[1mm]
$\bullet$ The dynamical content of the $TPEP$ is associated with three families of diagrams, 
shown in the figure below.
Diagrams of family I correspond to the minimal realization of chiral symmetry and involve 
only the $\p N$ coupling constants $g_A$ and $ f_\pi $. 
Family II describes effects associated with pion-pion correlations, whereas the
interactions in family III depend on the LECs, represented by black dots.

\vspace{2mm}

\begin{figure}[H]
\begin{center}
\includegraphics[width=0.80\columnwidth,angle=0]{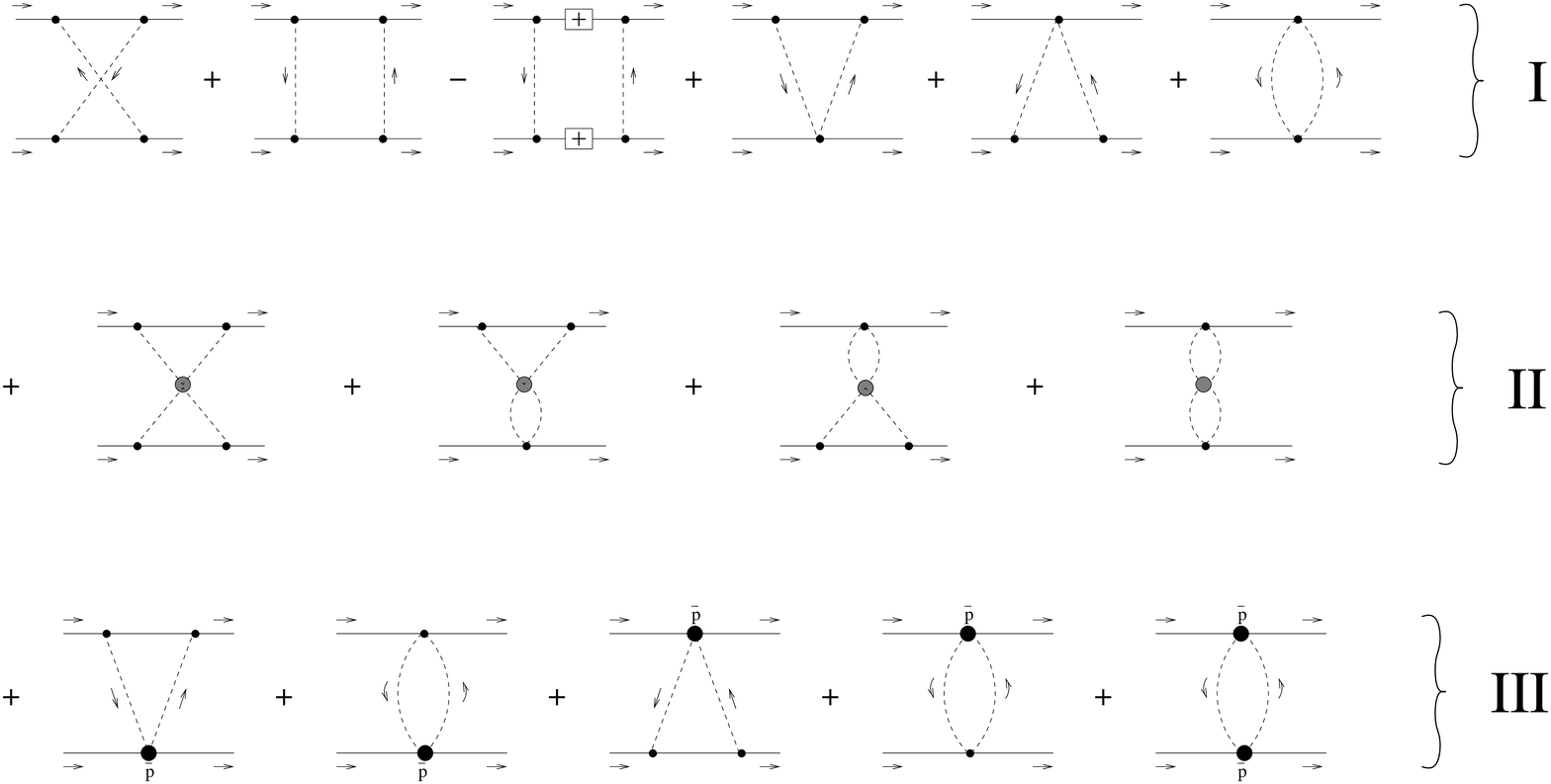}
\end{center}
\end{figure}

\vspace{-1mm}

\ni
$\bullet$ The relation of these diagrams with $\pi N$ scattering is well understood and may be used
to fix the (LECs) in family III.
When this procedure is adopted, the $TPEP$ does not contain any free parameters 
and becomes fully determined.\\[1mm]
$\bullet$ There is no room for scalar mesons, such as the $\s$, 
in the long and medium range parts of the $TPEP$. 


\ni
$\bullet$ Relativistic effects are visible in the final form of the $TPEP$ and arise 
from the proper covariant treatment of loop integrals.
Therefore they are present even when the external nucleon momenta are small. \\[1mm]
$\bullet$ Due to the treatment of loop integrals, heavy baryon and relativistic 
derivations of the potential do not coincide.
This problem is conceptually important, since it is related with the form of the asymptotic
chiral theorems.
From the point of view of internal theoretical consistency, the covariant procedure is favored.
On the other hand, heavy baryon calculations have the advantage of producing analytical results.
As far as numerical applications are concerned, the differences between both 
approaches are small.

\begin{figure}[H]
\begin{center}
\includegraphics[width=0.45\columnwidth,angle=0]{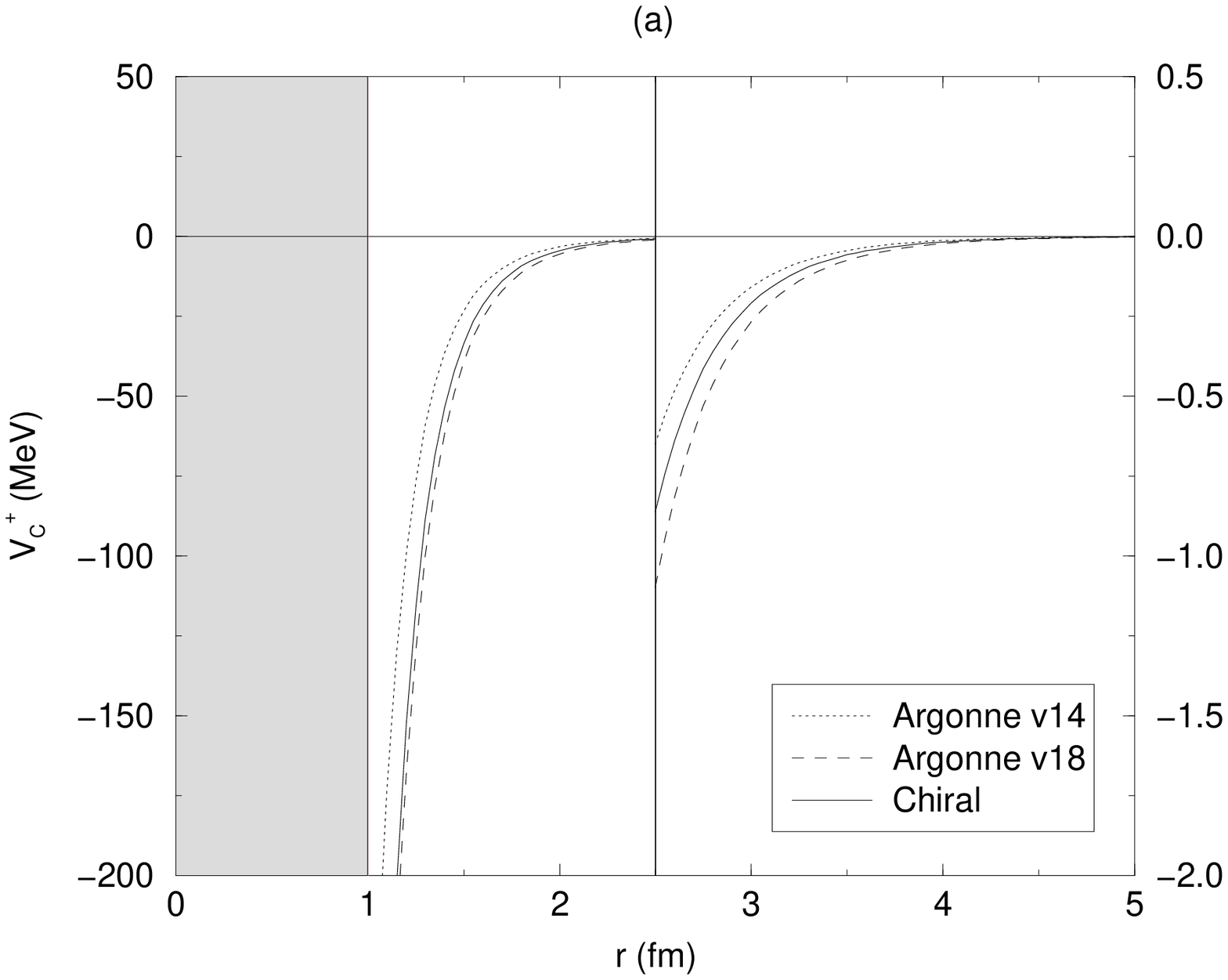}
\hspace{8mm}
\includegraphics[width=0.45\columnwidth,angle=0]{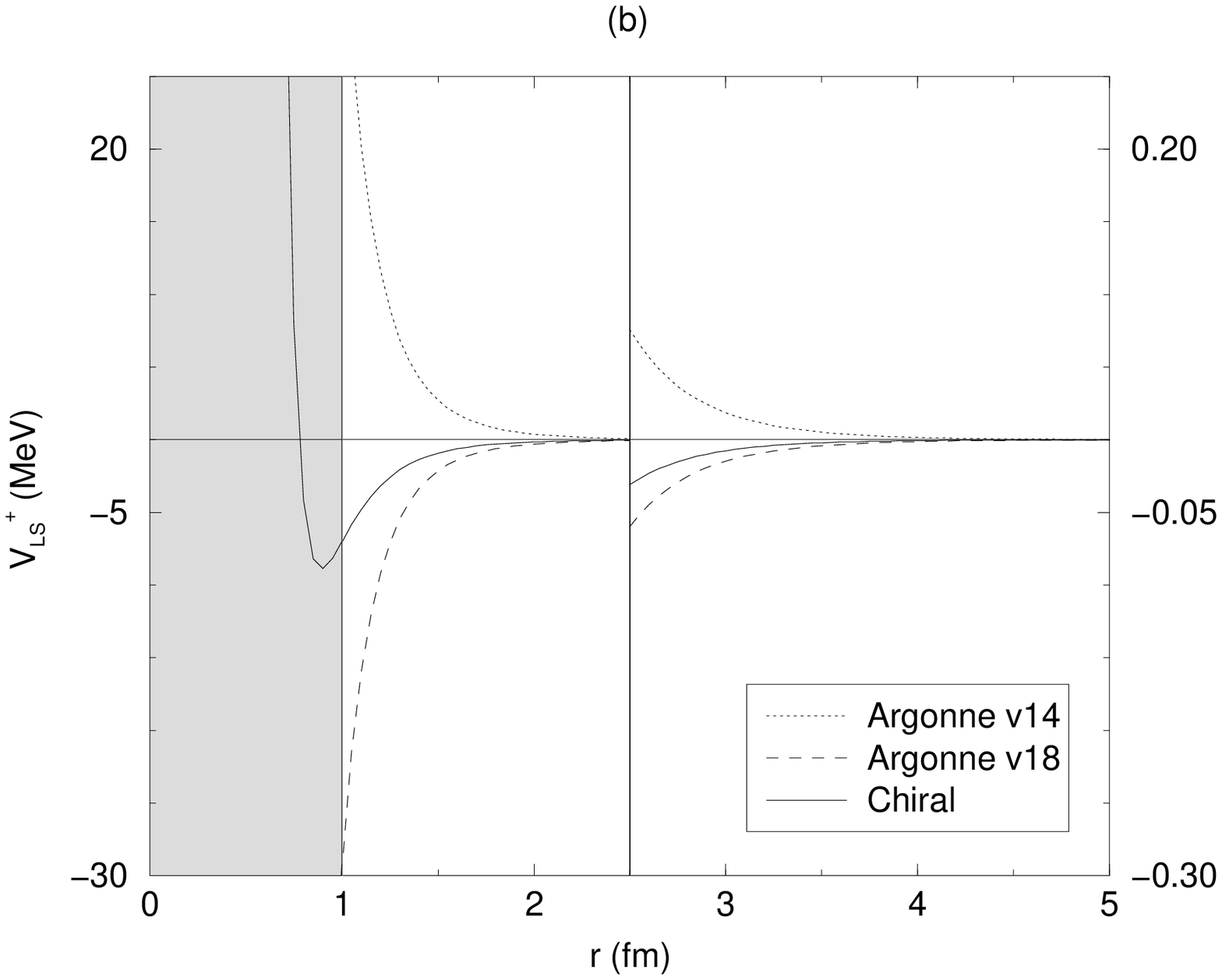}
\end{center}
\end{figure}




\ni
$\bullet$  The chiral picture is well supported by partial wave analyses\cite{OL}.
In the figure above we compare the chiral {TPEP} $V_C^+$ and $V_{LS}^+$ components with
results from the Argonne group\cite{A}. \\[1mm]
$\bullet$ The dominant features of the isospin independent component of the central potential 
are directly related with the QCD vacuum through the nucleon scalar form factor\cite{R}. \\[1mm]

\vspace{-2mm}

\begin{figure}[H]
\begin{center}
\includegraphics[width=0.45\columnwidth,angle=0]{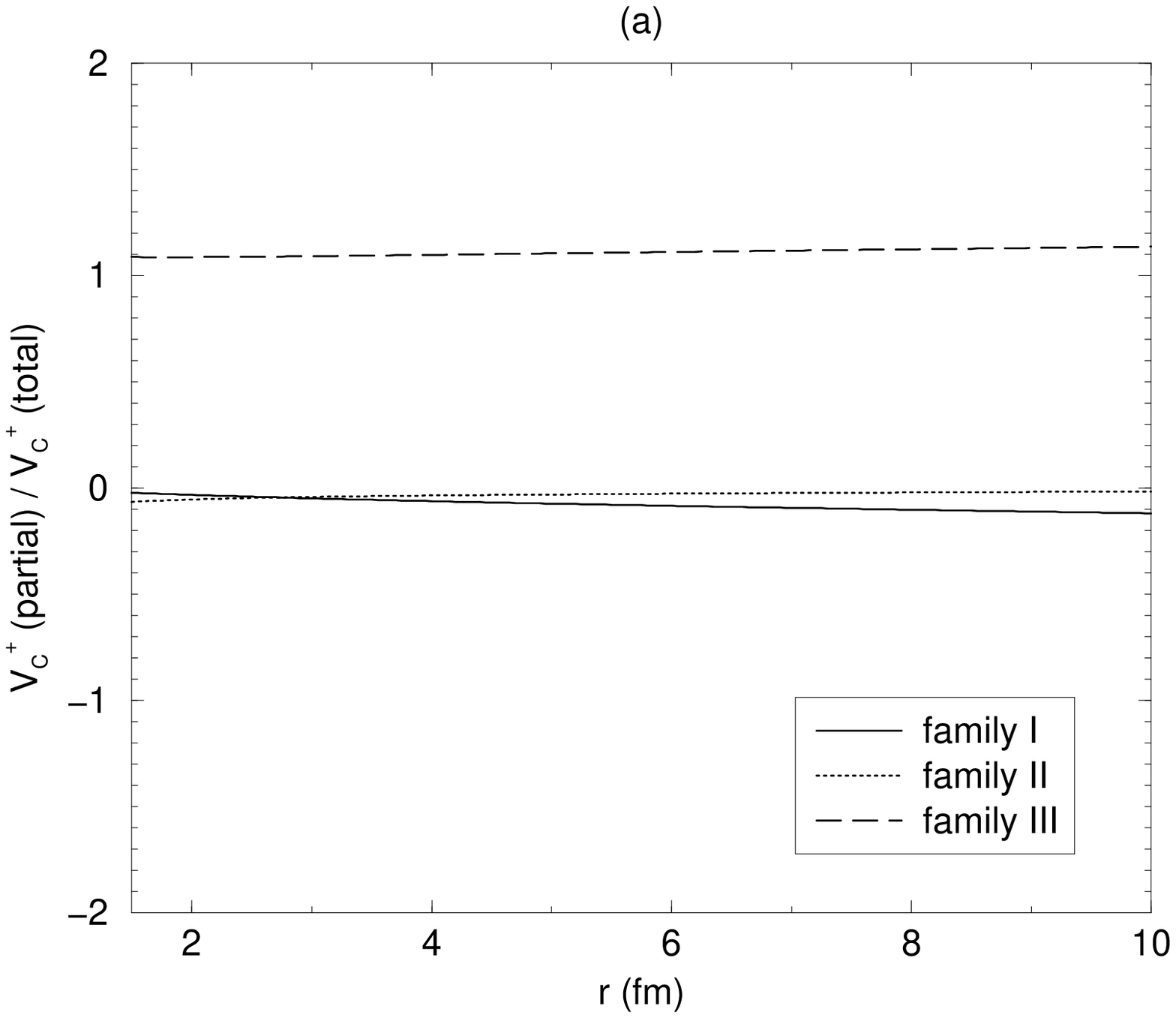}
\hspace{6mm}
\includegraphics[width=0.45\columnwidth,angle=0]{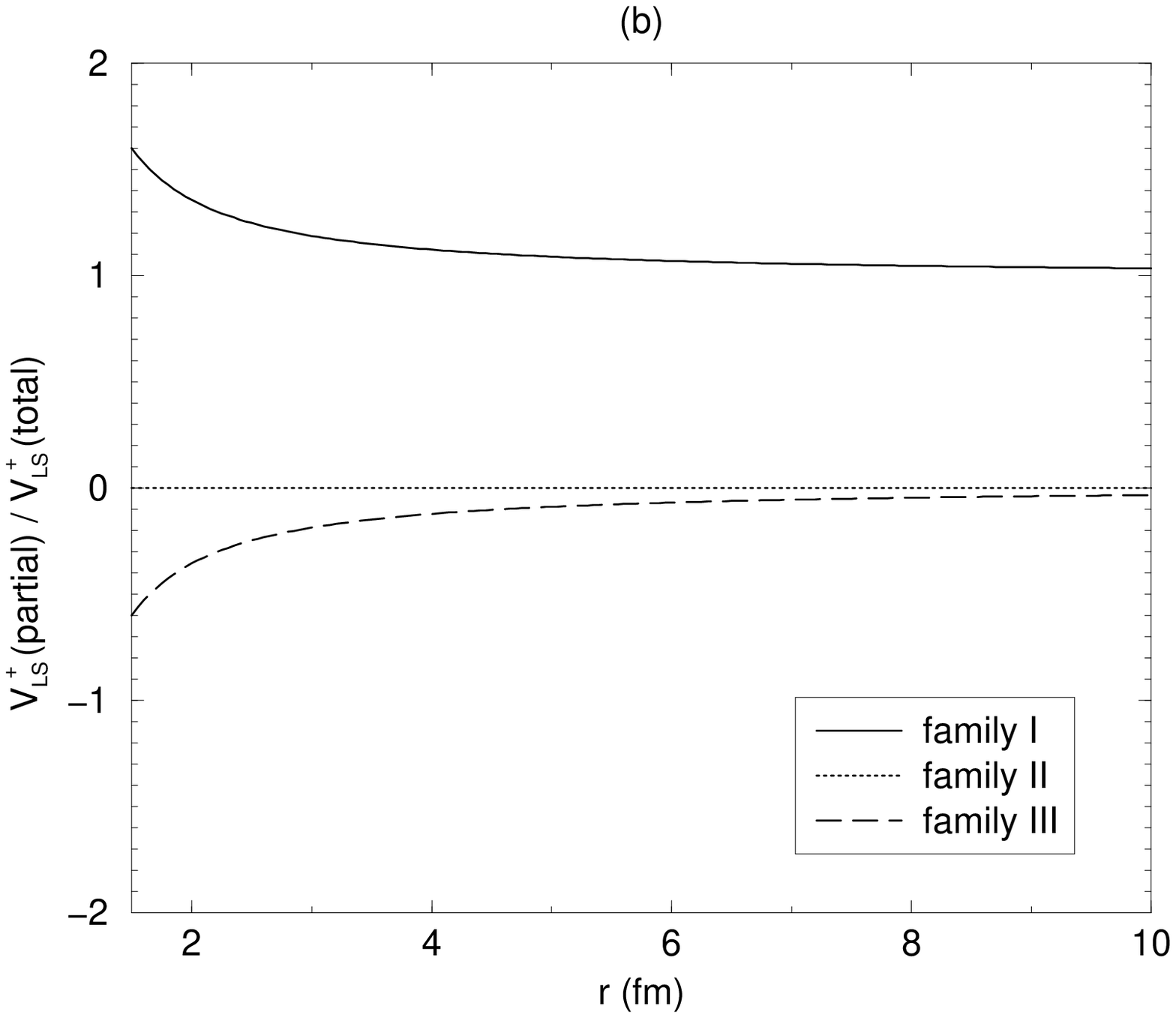}
\end{center}
\end{figure}




\ni
$\bullet$ As far as dynamics is concerned, the various channels of the potential 
are clearly dominated by isolated contributions arising from either family I or III and
intermediate $\pi\pi$ scattering processes in family II are almost completely irrelevant. 
This is in sharp contrast with older models for the $TPEP$, which were not based on chiral symmetry.
As typical instances, in the preceding figure we show the ratios of the individual contributions from 
families I, II and III by their sum, in the case of the components $V_C^+$ and $V_{LS}^+$. \\[1mm]
$\bullet$ The relative importances of $O(q^2)$, $O(q^3)$ and $O(q^4)$ terms in all the components 
of the potential has been assessed.
At distances of physical interest, they are consistent with converging series,
with the exception of the isospin independent central potential.
In the figure that follows we display the relative contribution of each chiral order 
to the $TPEP$ for $V_C^+$ and $V_{LS}^+$.
The black dots in the curves correspond to the points where the ratio is $0.5$.

\vspace{4mm}

\begin{figure}[H]
\begin{center}
\includegraphics[width=0.45\columnwidth,angle=0]{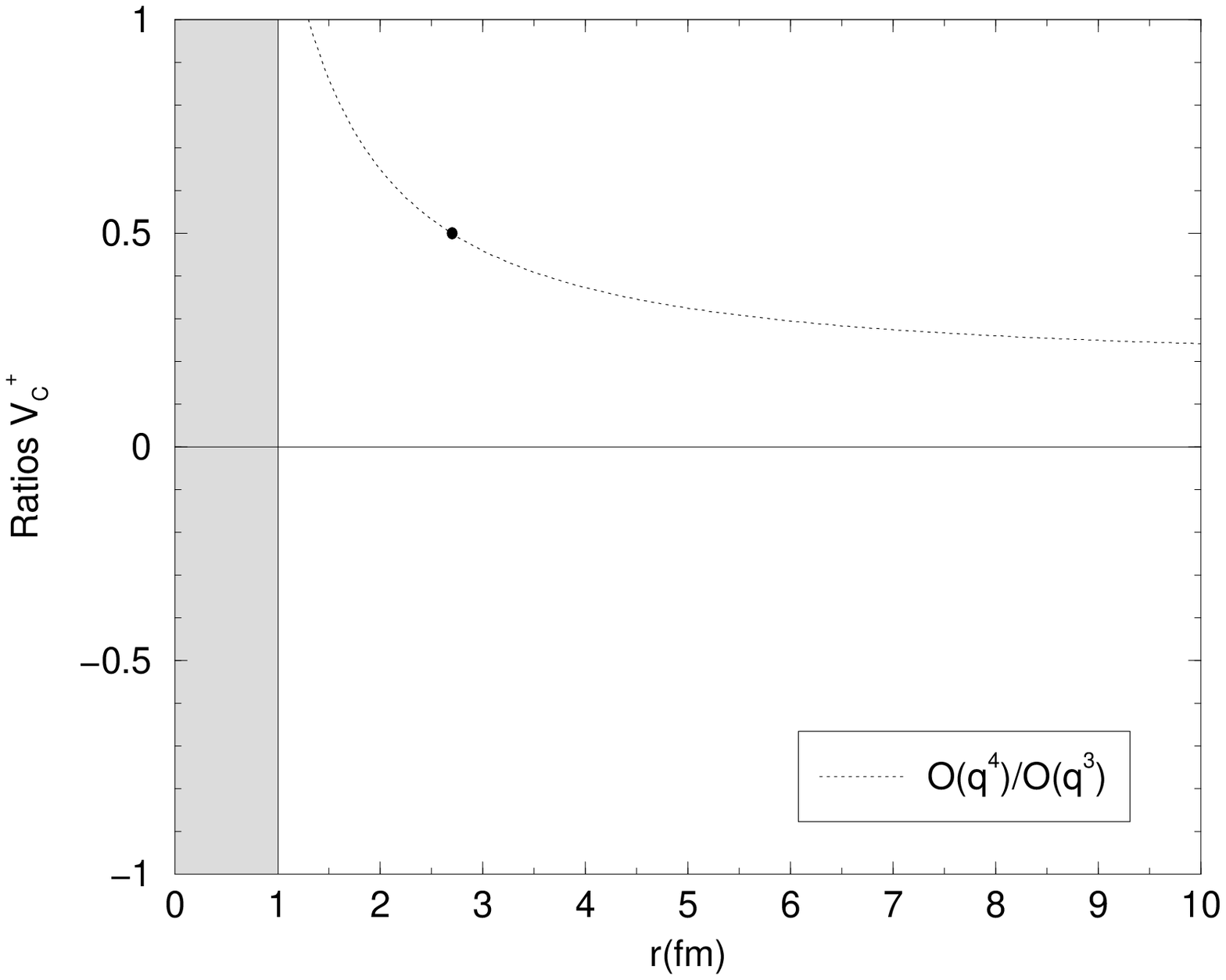}
\hspace{8mm}
\includegraphics[width=0.45\columnwidth,angle=0]{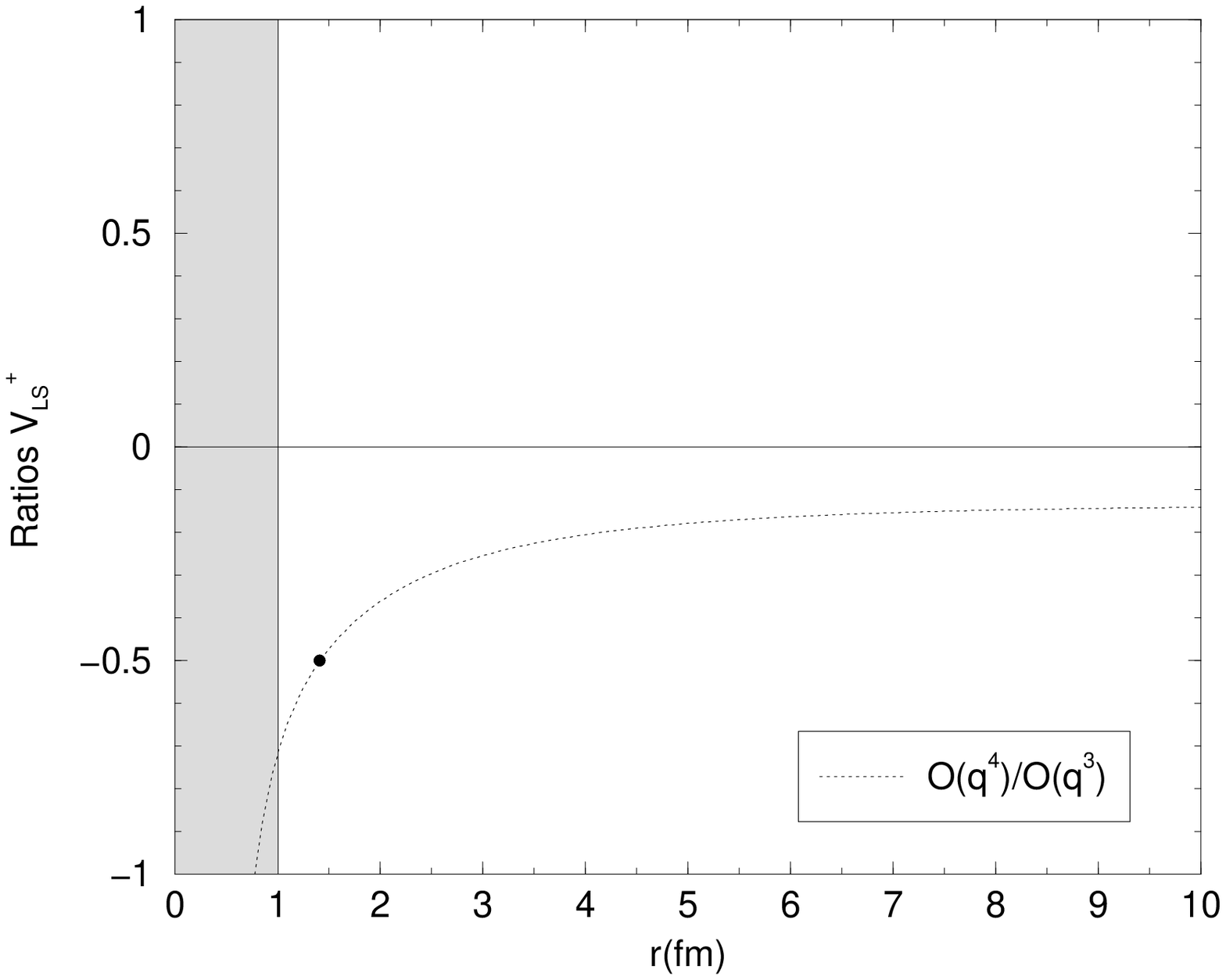}
\end{center}
\end{figure}




The clear picture of the $TPEP$ dynamics promoted by chiral symmetry allows one to identify
some problems that remain open and deserve being tackled in order to put the potential in a 
yet firmer basis. \\[1mm]
$\bullet$ The construction of the potential involves loop integrals, which must be regularized.
At present, the best regularization procedure for chiral symmetry in the baryon sector is 
the infrared scheme\cite{IR}, which gives rise to power counting.
Even if indications are that the influence of the regularization scheme is restricted to 
distances smaller than 1 fm, the $TPEP$ problem remains in the want of a full calculation 
based on the infrared method.
\\[1mm]
$\bullet$ A rather puzzling aspect of the chiral $TPEP$ is that its leading terms are 
formally predicted to be $O(q^2)$, whereas the all important central isospin independent
component $V_C^+$ begins at $O(q^3)$. 
This may be associated with the poor convergence of the chiral series for this term, 
as shown in the preceding figure.  
The numerical reasons for its odd behavior can be traced back to the large size 
of the LECs that are used in the first two diagrams of family III.
These LECs, in turn, are dynamically generated by processes involving delta intermediate states.
Therefore the explicit inclusion of delta degrees of freedom in a covariant calculation could 
prove useful in shedding light into this problem. \\[1mm]
$\bullet$ The relation of the potential with data is very important.
At present, one has good indications that the chiral $TPEP$ is able to reproduce well
empirical phase shifts.
However, a problem that occurs in this kind of testing is that the theoretical potential 
can only be directly used in the study of peripheral waves, which are small and carry 
large uncertainties.  
In order to study a larger set of waves and energies, theoretical expressions have to 
be corrected at short distances by means of cutoffs or form factors, which also influence
numerical results.
As this problem cannot be avoided, a proper assessment of the merits of the $TPEP$ could 
be obtained by mapping in detail the influence of cutoffs and form factors over numerical 
results. \\[1mm] 
$\bullet$ A related problem concerns the determination of the numerical values of the LECs
present in the effective lagrangians. 
Many of the LECs relevant to the $NN$ interaction also contribute to elastic $\p N$ scattering
and it would be useful to know whether values extracted from these two processes are compatible.
In doing this comparison, it is important to bear in mind that the numerical values for 
the LECs depend on the chiral order of the expansion one is working with. \\[1mm]
$\bullet$ In the long run, it would be interesting to consider the extension of the chiral picture 
to potentials used in many-body calculations which, for technical reasons, tend to be more schematic.
Usually, they rely heavily on scalar-isoscalar interactions inspired in the linear $\s$ model.
However, the chiral $TPEP$ does not support this assumption, especially as far as the $O(q^3)$
nature of the central potential is concerned.


\begin{thebibliography}{0}

\bibitem{TNS} M. Taketani, S. Nakamura and M. Sasaki, Progr. Theor. Phys. {\bf VI}, 581 (1951).

\bibitem{EE} E. Epelbaum, preprint nucl-th/0509032.

\bibitem{HRR} R. Higa, M.R. Robilotta and C. A. da Rocha, Phys. Rev. C {\bf 69}, 034009 (2004);
R. Higa and M.R. Robilotta, Phys. Rev. C {\bf 68}, 024004 (2003);
J-L. Ballot, M. R. Robilotta and C. A. da Rocha, Phys. Rev. C {\bf 57}, 1574 (1998);
M. R. Robilotta an C. A. da Rocha, Nucl. Phys. A {\bf 615}, 391 (1997);
M. R. Robilotta, Nucl. Phys. A {\bf 595}, 171 (1995);
C. A. da Rocha and M. R. Robilotta, Phys. Rev. C {\bf49}, 1818 (1994).

\bibitem{WNN} S. Weinberg, Phys. Lett. B {\bf 251}, 288 (1990); 
Nucl. Phys. B {\bf 363}, 3 (1991).

\bibitem{HB} N. Kaiser, R. Brockman and W. Weise, Nucl. Phys. A {\bf 625}, 758 (1997);
N. Kaiser, Phys. Rev. C{\bf 64}, 057001 (2001);
N. Kaiser, Phys. Rev. C{\bf 65}, 017001 (2001);
E. Epelbaum, W. Gl\"ockle and Ulf-G. Meissner, Nucl. Phys. A {\bf 637}, 107 (1998); 
Nucl. Phys. A {\bf 671}, 295 (2000);
D.R. Entem and R. Machleidt, Phys. REv. C {\bf 66}, 014002 (2002).

\bibitem{OL} http://nn-online.org; http://gwdac.phys.gwu.edu.

\bibitem{A} R. B. Wiringa, R. A. Smith and T. L. Ainsworth, Phys. Rev. C {\bf 29}, 1207 (1984);
R. B. Wiringa, V. G. J. Stocks, and R. Schiavilla, Phys. Rev. C {\bf 51}, 38 (1995).
  
\bibitem{R} M. R. Robilotta, Phys. Rev. C {\bf 63}, 044004 (2001).

\bibitem{IR} P. J. Ellis and H-B. Tang, Phys. Rev. C {\bf 57}, 3356 (1998);
K. Torikoshi and P. Ellis, Phys. Rev. C {\bf 67}, 015208 (2003);
T. Becher and H. Leutwyler, Eur. Phys. J. C {\bf 9}, 643 (1999);
J. High Energy Phys. {\bf 106}, 17 (2001).



\end{thebibliography}
\end{document}